\documentclass[longauth]{aa}

\usepackage{natbib,twoopt}
\usepackage[breaklinks=true]{hyperref} 
\bibpunct{(}{)}{;}{a}{}{,} 
\makeatletter
\newcommandtwoopt{\citeads}[3][][]{\href{http://adsabs.harvard.edu/abs/#3}%
{\def\hyper@linkstart##1##2{}%
\let\hyper@linkend\@empty\citealp[#1][#2]{#3}}}
\newcommandtwoopt{\citepads}[3][][]{\href{http://adsabs.harvard.edu/abs/#3}
{\def\hyper@linkstart##1##2{}%
\let\hyper@linkend\@empty\citep[#1][#2]{#3}}}
\newcommandtwoopt{\citetads}[3][][]{\href{http://adsabs.harvard.edu/abs/#3}%
{\def\hyper@linkstart##1##2{}%
\let\hyper@linkend\@empty\citet[#1][#2]{#3}}}
\newcommandtwoopt{\citeyearads}[3][][]%
{\href{http://adsabs.harvard.edu/abs/#3}
{\def\hyper@linkstart##1##2{}%
\let\hyper@linkend\@empty\citeyear[#1][#2]{#3}}}
\makeatother

\usepackage[varg]{txfonts}

\begin{document}

\title{EMIR, the near-infrared camera and multi-object spectrograph for the GTC}
\subtitle{EMIR@GTC} 

\author{F. Garz\'on \inst{\ref{inst1},\ref{inst2}} 
        \and M. Balcells \inst{\ref{inst3},\ref{inst1},\ref{inst2}}
        \and J. Gallego \inst{\ref{inst4}}
        \and C. Gry \inst{\ref{inst5}}
        \and R. Guzm\'an \inst{\ref{inst7}}
        \and P. Hammersley \inst{\ref{inst6}}
        \and A. Herrero\inst{\ref{inst1},\ref{inst2}}
        \and C. Mu\~noz-Tu\~n\'on \inst{\ref{inst1},\ref{inst2}}
        \and R. Pelló \inst{\ref{inst5}}
        \and M. Prieto\inst{\ref{inst1},\ref{inst2}}
    \and É. Bourrec \inst{\ref{inst9}}
        \and C. Cabello\inst{\ref{inst4}}
        \and N. Cardiel \inst{\ref{inst4}}
        \and C. González--Fernández \inst{\ref{inst13}}
        \and N. Laporte  \inst{\ref{inst8}}
        \and B. Milliard \inst{\ref{inst5}}
        \and S. Pascual \inst{\ref{inst4}}
        \and L.~R. Patrick  \inst{\ref{inst10},\ref{inst11}}
        \and J. Patrón  \inst{\ref{inst1},\ref{inst2}}
        \and S. Ramírez--Alegría  \inst{\ref{inst12}}
        \and A. Streblyanska  \inst{\ref{inst1},\ref{inst2}}
}

\institute{Instituto de Astrof\'{\i}sica de Canarias, IAC, V\'{\i}a Láctea s/n, 38205, La Laguna (S.C. Tenerife), Spain \email{fgl@iac.es}\label{inst1}
   \and Departamento de Astrof\'{\i}sica, Universidad de La Laguna, 38206, La Laguna (S.C. Tenerife), Spain\label{inst2}
   \and Isaac Newton Group of Telescopes, ING, 38700 La Palma (S.C. Tenerife), Spain\label{inst3}
   \and Departamento de Física de la Tierra y Astrofísica \& Instituto de Física de Partículas y del Cosmos (IPARCOS). Universidad Complutense de Madrid. 38206, Madrid, Spain\label{inst4}
   \and Aix Marseille Univ, CNRS, CNES, LAM, Marseille, France \label{inst5}
   \and Dept. of Astronomy. University of Florida, USA\label{inst7}
   \and European Southern Observatory, ESO, Garching bei München, Germany\label{inst6}
   \and Institut de Recherche en Astrophysique et Planétologie (IRAP), Université de Toulouse, CNRS, UPS, CNES, 14 Av. Edouard Belin, 31400, Toulouse, France \label{inst9}
   \and Institute of Astronomy, University of Cambridge, Madingley Rise, Cambridge CB3 0HA, UK \label{inst13}
   \and Kavli Institute for Cosmology, University of Cambridge, Madingley Road, Cambridge, CB3 0HA, UK \label{inst8}
   \and Departamento de Física Aplicada, Universidad de Alicante, San Vicente del Raspeig, E-03690 Alicante, Spain \label{inst10}
   \and School of Physical Sciences, The Open University, Walton Hall, Milton Keynes MK7 6AA, UK \label{inst11}
   \and Centro de Astronomía (CITEVA), Universidad de Antofagasta, Av. Angamos 601, Antofagasta, 1271155, Chile \label{inst12}
}

\abstract {

We present EMIR, a powerful near-infrared (NIR) camera and multi-object spectrograph (MOS) installed at the Nasmyth focus of the 10.4 m GTC. EMIR was commissioned in mid-2016 and is offered as a common-user instrument. It provides 
spectral coverage of 0.9 to 2.5 $\mu m$ over a field of view (FOV) of $6.67\arcmin\times6.67\arcmin$ in imaging mode, and $6.67\arcmin\times4\arcmin$ in spectroscopy. EMIR delivers up to 53 spectra of different objects thanks to a robotic configurable cold slit mask system that is located inside the cryogenic chamber, allowing rapid reconfiguration of the observing mask. The imaging mode is attained by moving all bars outside the FOV and then leaving an empty space in the GTC focal surface. The dispersing suite holds three large pseudo-grisms, formed by the combination of high-efficiency FuSi ion-etched ruled transmission grating sandwiched between two identical ZnSe prisms, plus one standard replicated grism. These dispersing units offer the spectral recording of an atmospheric window $J,H,K$ in a single shot with resolving powers of 5000, 4250,  4000, respectively for a nominal slit width of 0.6\arcsec, plus the combined bands $YJ$ or $HK$, also in a single shot, with resolution of $\sim$ 1000. The original Hawaii2 FPA detector, which is prone to instabilities that add noise to the signal, is being replaced by a new Hawaii2RG detector array, and is currently being tested at the IAC. 
This paper presents the most salient features of the instrument, with emphasis on its observing capabilities and the functionality of the configurable slit unit. Sample early science data is also shown.  }

\keywords{Astronomical instrumentation, methods and techniques -- instrumentation: spectrographs -- techniques: spectroscopic -- techniques: image processing}

\titlerunning{EMIR@GTC}
\authorrunning{F. Garz\'on and EMIR Consortium} 

\maketitle

\section{Introduction.}
Modern astrophysics cannot be understood without the constant development of more powerful
telescopes and instrumentation that make it possible to observe increasingly distant objects or that
significantly improve the body of observational data available for already known objects. This is achieved through
the improvement in spatial resolution and sensitivity and novel observing modes that new instruments
offer in combination with ever-larger telescopes,  thus providing better image quality. After the opening in the last decades of a  suite of 8--10m class telescopes, which will soon be complemented with the Vera Rubin Telescope, a great deal of effort is now being concentrated in the new class of extremely large telescopes, which can reach up to 40m in diameter, and in the commissioning  of the JWST,  now in operation after its successful launch in December 2021.

All these large telescopes are equipped with a suite of specially designed instruments to make the most of the enormous observational possibilities they offer, and which translate into an increase of several orders of magnitude in light collection capacity and a substantial improvement in spatial resolution, as well as superior image quality compared to the currently existing 4m class telescopes. These instruments, already in use or under development, cover a wide range of the electromagnetic spectra accessible from Earth and offer varied observation possibilities, both in images and in spectroscopy. The design and construction of these instruments represents a formidable technological challenge, motivated by the need to maximise the performance of large telescopes, which imposes very large geometric sizes of the opto-mechanical elements, with stringent tolerances in their mounting and alignment, and qualities in the surface finishing of the optics and in the engineering of the mechanics that require careful design and development prior to  manufacture. Additionally, these instruments are equipped with large-format detector mosaics, which implies the development of novel control electronics capable of conducting enormous amounts of data   at increasingly higher speeds. The large amount of data from each individual reading, combined with the speed of  acquisition and the exotic nature of the observation modes, requires the development of automatic algorithms for the treatment and reduction of these data, which offer the astronomer data products devoid of instrumental effects and ready for analysis.

\begin{figure}
        \begin{center}
                \includegraphics[width=\columnwidth]{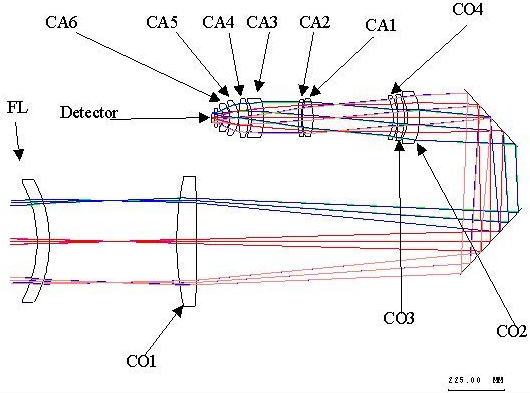}
                \caption{Optical layout of EMIR. The labels correspond to: field lens and cryostat window (FL), collimator lenses 1 to 4 (CO\#), and  camera lenses 1 to 6 (CA\#) }
        \label{folayout}
        \end{center}
\end{figure}

\begin{figure}
        \begin{center}
                \includegraphics[width=\columnwidth]{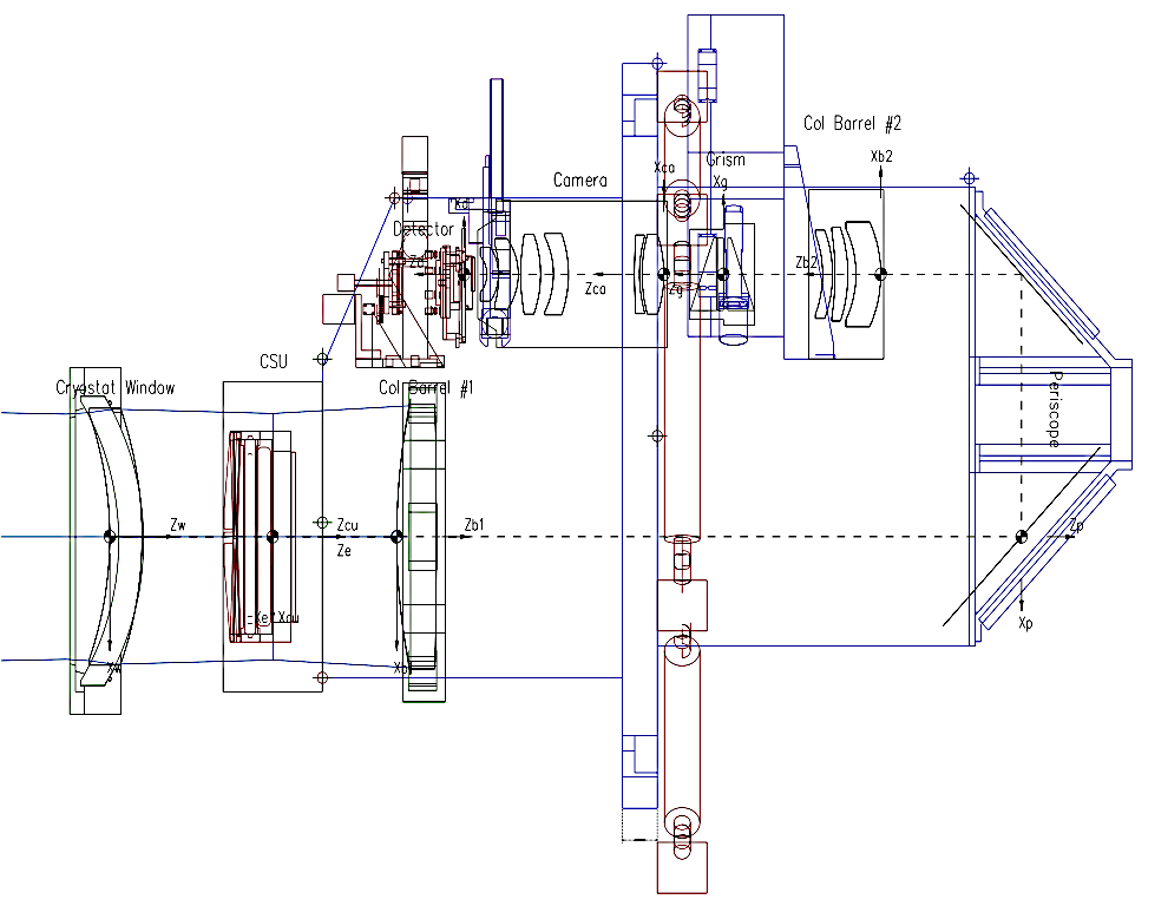}
                \caption{Opto-mechanical layout of EMIR. The mounting of the different units and mechanism in the optical bench is shown. See references in text for details.}
        \label{fomlayout}
        \end{center}
\end{figure}

Among the large instrumental suite in each of these new telescope stations, multi-object
spectrographs (MOS)  play a prominent role and are always in the top rank of the most demanded instruments. The near-infrared (NIR) MOS is particularly interesting for large telescopes as the NIR permits the detection of objects  in the optical bands hidden by massive material clouds, and because of the effect of redshift in objects at cosmological distances. Given the next and upcoming generation of ground- and space-based observatories (e.g. JWST, E-ELT), NIR spectroscopy is of increasing importance. To support 30m class telescopes, well-established NIR spectrographs on 8--10m class telescopes are a vital resource, and are scientifically viable in their own right. EMIR (\citeads{2016ASPC..507..297G}, \citeads{2006SPIE.6269E..18G, 2016SPIE.9908E..1JG, 2017hsa9.conf..652G}) is occupying this niche  at the Gran Telescopio Canarias  (GTC). It is one of the common user instruments at the GTC, and offers observing capabilities in imaging and spectroscopy in both long-slit  and multi-object modes, over a field of view of $6.67\arcmin\times6.67\arcmin$ in image mode and $6.67\arcmin\times4\arcmin$ in spectroscopy. It had its first light at the GTC in mid-2016  and was initially offered to the community in late 2017 after a long period of commissioning and science verification phases. The control system of EMIR was constantly tuned during the first years to permit a smooth coordination with the telescope and proper functioning of the different observing modes. EMIR was tested in the lab with a Hawaii2 FPA detector of the initial architecture, but we were forced to replace it, rather close to the shipping date to the GTC, due to the high risk of explosion that had happened in other similar arrays. The new detector, while free from the risk of explosion thanks to the new manufacturing process, was less adequate than the original one, due to higher read-out noise and instabilities in the signal. All together, these features led to an overall instrument performance below  what  was expected, in particular for faint sources.

This unforeseen problem forced the instrument team to devote substantial efforts to try to alleviate the lower effective sensitivity. While in image mode we   developed methods to clean the frame with remarkable success, in spectroscopy only minor improvements were achieved despite the many attempts using different approaches. It should be noted that this caveat is of particular importance, but only for faint objects. Finally, new funds arrived that permit the replacement of the detector by a new Hawaii2RG, which is currently being tested at the IAC. This was the motivation for   deferring   the publication of the  paper; we now feel, given that  the  replacement detector  will be ready in the near future,  that it is time to have a  reference in the literature for this instrument from the astronomical standpoint.

This paper is organised as follows. In section \ref{sec:emir} a brief description of the EMIR instrument is given that includes several references to technical papers where the interested reader can find a more in-depth description of the different  subunits. Sections \ref{sec:csu} and \ref{sec:cat} outline the process of multi-slit mask design. Section \ref{sec:pyemir} introduces the instrument data reduction pipeline. Section \ref{sec:sci} presents some results obtained during the commissioning and science verification phases. Finally, section \ref{sec:conclu} summarises the main features of the EMIR instrument and its observing capabilities.

\section{EMIR instrument.}
\label{sec:emir}
EMIR is a NIR wide-field spectrograph, similar to other instruments developed in the 2000s for 10m class telescopes, such as KMOS/VLT\footnote{\url{http://www.eso.org/public/teles-instr/paranal-observatory/vlt/vlt-instr/kmos}}, MOSFIRE/Keck\footnote{\url{https://www2.keck.hawaii.edu/inst/mosfire/home.html}}, and Flamingos2/GEMINI\footnote{\url{https://www.gemini.edu/instrumentation/flamingos-2}}. The concept of the instrument corresponds to a classical  camera and spectrograph design, but with several novelties. Its spectral working range is dictated by the spectral response of the detector, hence it spans from 0.9 to 2.5 $\mu m$. From the beginning the top priority mode of EMIR was set as multi-object spectroscopy in the $K$ band, due to the requirements of the instrument main scientific driver, the Galaxy Origins and Young Assembly (GOYA) Survey (\citeads{2003RMxAC..16..209G}, \citeads{2003RMxAC..16...69B}). This imposed a fully cryogenic layout of the instrument that, in addition, set a severe restriction for the multi-object capability of EMIR.

EMIR has two main observing modes: imaging and spectroscopy. The image field of view subtends 6.67\arcmin square, while the FOV available for placing slits in spectroscopic mode is restricted to the central 4\arcmin\ in one direction, aligned with the sliding bars of the Cold Slit Unit (CSU; see below) and covers the full field in the direction perpendicular to it.

The EMIR optical and opto-mechanical layouts can be seen in figs. \ref{folayout}  and \ref{fomlayout}. The cryostat window acts as a field lens that increases the GTC focal ratio by approximately 10\%, resulting in a  f/17.7 in the modified focal plane of the telescope, which sits inside the dewar, right after the cryostat window. The purpose of this increase in the focal ratio is to diminish the intrinsic curvature of the surface plane as this is the location of the CSU, a cryogenic robotic system that provides EMIR with the capability of configuring multi-slit patterns.

\begin{figure}
        \begin{center}
                \includegraphics[width=\columnwidth]{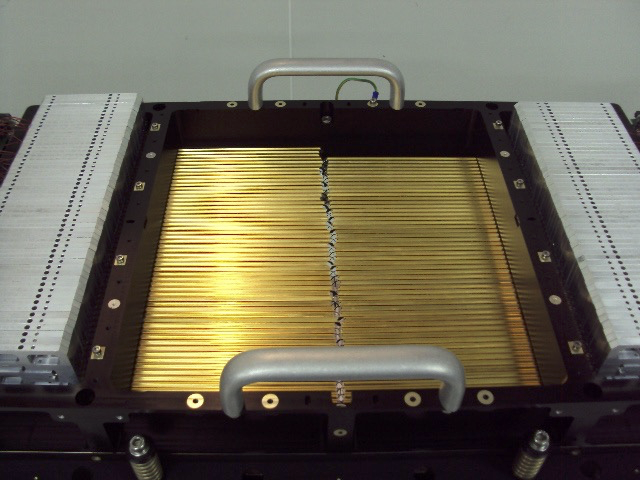}
                \caption{CSU of EMIR prior to being mounted on the EMIR optical bench. The 55 pairs of sliding bars, gold covered, can be clearly seen.}
        \label{fcsu}
        \end{center}
\end{figure}

The CSU \citepads{2012SPIE.8446E..5NT} shown in fig. \ref{fcsu}  is a distinct feature of EMIR. It is a large mechanical component which is attached to the EMIR optical bench and located at the GTC focal plane. It holds 55 pairs of sliding bars that can be positioned with very high accuracy in the central area of the mechanical frame. Each pair can form a slit of selectable width in a specific area of the FOV. Several slits can be jointly aligned along the same direction to form a longer slit and the full set of 55 slits can be aligned together to form a classical long slit. Each slit is equipped with a custom designed actuator based on a piezo stick-slip concept that provides motion in both directions, plus a capacitive position sensor (see details in \citeads{2012SPIE.8446E..5NT}). The image mode is attained by moving all bars outside the FOV then leaving an empty space in the GTC focal surface. The CSU is in essence a 2D component of large physical dimensions (470 mm of side length in the frame FOV), hence the need to improve the curvature of the focal surface to keep the image quality figure within the prescribed margins. As a result of the geometrical distortion of the EMIR camera, the two end slits are partially vignetted so that, for practical purposes, the number of usable slits is limited to 53.

\begin{figure}
        \begin{center}
                \includegraphics[width=\columnwidth]{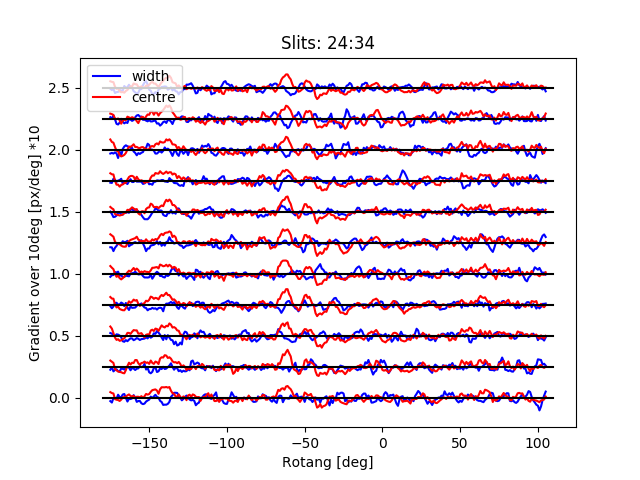}
                \caption{Displacement of slit centres (red lines) and slit widths (blue lines) with the turn of EMIR in the Nasmyth rotator for the slit numbers given in the plot title. These are displayed in terms of gradients of the magnitudes per 10\degr\ of rotation angle. The values are multiplied by a factor of 10 to visualise them. See text for explanation.}
        \label{fcsu_stab}
        \end{center}
\end{figure}

The opposite bars, each one of 370 mm in length, can be positioned with a mean accuracy better than 6 $\mu m$, for a typical slit geometrical width in the range between 500 and 1000 $\mu m$, hence ensuring the constancy of the slit width, and so the spectral resolution limit, for the case of multi-bar long slits. The plate scale at the focal plane of EMIR, where the CSU is located, is 1.1717 \arcsec/mm;  combined with the f/1.91 focal ratio of the camera, it yields a pixel size of 0.1947\arcsec\ projected on the sky and 166 $\mu m$ in the CSU plane. Hence, standard slit widths measure roughly from 3 to 6 px, or 0.6\arcsec\ to 1.2\arcsec, while narrower or wider slits are also attainable. The gold coating on the exposed side of the bars rejects over 98\% of the incoming flux, which prevents the mechanical stretching of the bars due to heating. The mean displacement of bars due to expansion by heat has been measured by detecting the bar edge on the detector image and has resulted to be less than 0.015 px/hour, or 0.003 \arcsec/hour in the sky. This   permits the use of long exposures times without the need to reconfigure the CSU. The mechanical stability of the unit also plays an important role when pointing the instrument at faint objects that require long observing times. In fig. \ref{fcsu_stab} we show the results of a test measurement on which the 55 slits are positioned at different, but fixed, locations in the focal plane. The location of each bar is measured at the detector every degree while the instrument rotates a full circle. Figure \ref{fcsu_stab} displays the gradients of slit centres and widths per 10\degr\ of rotation angle  with respect to the central rotator position of each gradient measurement. Each slit result is shifted by 0.25 px in Y and a black solid line indicates the zero value of each one, which are plotted in pixel units using the Y-axis scale. Around rotation angles of -55\degr\ and -145\degr\ in fig. \ref{fcsu_stab}  there are noticeable increases in the gradients of both slit widths and centres. These bumps are due to discontinuities in the rotator tracks that provoke jumps in the instrument position. The maximum absolute values of the gradients in  fig. \ref{fcsu_stab} are $1.03\times10^{-2}\pm2.7\times10^{-3}$ and $1.36\times10^{-2}\pm3.6\times10^{-3}$ for the gradients in slit widths and centres, respectively, including the bumps. The net effect in the observations is negligible as these figures translate to a maximum variation in slit width of 0.002\arcsec and 0.003\arcsec\ in slit centre, both during a rotation of the instrument of 10\degr, which is a typical rotation range during one hour of integration time. The displacement figures are a tiny bit over the accuracy of the bar position measurement at the detector, hence the CSU is virtually rigid with respect to instrument rotation.

The width of each slitlet can be configured with almost no limit within the instrument geometry, but for practical and security reasons the narrowest permitted slit figure corresponds to one detector pixel, or slightly below 0.2\arcsec. The configuration of a multi-slit pattern of the CSU, on the other hand, is a task that requires some effort from the astronomer and the use of several software tools that are specifically designed for this purpose. This process, and its several tricks, are   described in section \ref{sec:csu}. For long-slit observations, the instrument control software has a predefined set of widths and positions of the slit in the FOV that can satisfy most of the user demands. It is always possible to tailor a long slit suited to the user's needs by making use of the CSU configuration tool and combining all the slits in a single alignment.

Right behind the CSU, the light beam hits the first lens of the collimator. The collimator unit is composed by a large single lens plus a triplet of lenses that are mounted after the periscope, a double flat mirror unit that folds the beam to compact the mechanical design of the instrument. 

Once the incoming beam is collimated, it enters into the spectral dispersive area of EMIR. The grism wheel supports three large barrels, each one composed of a high-quality fused silica transmission grating sandwiched between two twin ZnSe prisms. Each triplex prism--grating--prism is designed to function like a dispersing grism in one of the three atmospheric windows $J,H,K$. The three gratings are holographically etched to provide extremely regular groove densities of respectively 1032, 683, and 486 g.mm$^{-1}$ for $J, H,$ and $K$. The prisms and the grating are mounted with an innovative design with springs securing an extremely stable isostatic support that avoids any displacement during the temperature decrease in the cryostat. They offer a resolving power R of $\sim 5\,000$, $4\,250$, and $4\,000$ for $J, H$, and $K$, respectively. The dispersive suite is complemented by a standard grism that can cover either the combined $HK$ or $YJ$ spectral range in first and second order, respectively. In addition, the grism wheel also contains a cold Lyot stop for image mode.

After crossing the grism wheel, the beam is focused by a six-lens element camera that delivers an f/1.91 converging beam to the filter wheel, that includes a set of broad- and narrowband filters, and then to the detector array. The detector is mounted onto the detector translation unit (DTU). This is a 3D translation table on which the detector is mounted. Its motion permits the refocusing of the detector when needed by moving along the optical axis, Z, of the instrument, but it also works by displacing the detector on its proper plane (XY) (i.e. perpendicular to the optical axis), so that the internal flexures of the instrument during the rotation to track a sky position are corrected. The DTU has submicron accuracy in the XY motion and we have plans to exploit this extremely precise mechanism for adding capabilities to the instrument by displacing the detector in subpixel steps in the sequence of frames acquired during the integration to increase both the spatial and spectral resolution of the data.

The instrument and its many details of use are fully described on its website at http://research.iac.es/proyecto/emir/.

\begin{figure}
        \begin{center}
                \includegraphics[width=\columnwidth]{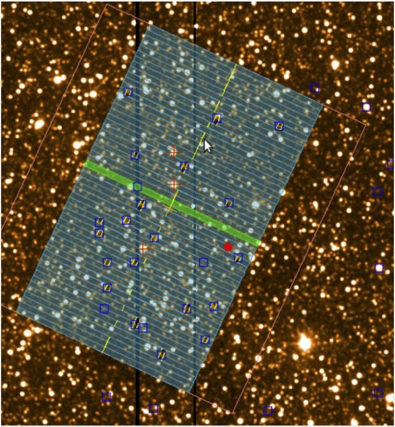}
                \caption{Snapshot of the main OSP panel showing the location of the mask superimposed on a sky image and with several slits allocated to targets (blue squares) and fiducial objects (red crosses). The background image is a UKIDSS field located in the Galactic plane at 30\degr\ longitude. This example was used to construct a mask for the GALEP project (see Sect. \ref{sec:cat}), where the targets are selected in the infrared colour--magnitude diagram as candidates for red giant stars.}
        \label{fosp}
        \end{center}
\end{figure}

\section{Observing strategy}

In spectroscopic NIR observations, it is common practice to use a nodding or beam switching scheme to facilitate the measure and removal of the sky background, which is quite often orders of magnitude brighter than the object.
The nodding throw in relation with a single slit length, $\sim$7.2\arcsec, has to be taken into account
whilst planning the observing strategy.
The presence of companion objects in the slit that can fall in the same or neighbours pixels as the object of interest, either in the initial pointing or in the nodded position, also deserves   consideration at the time of the mask design (see sec. \ref{sec:csu}).

Similarly, a novel aspect of NIR spectroscopic observations is telluric contamination.
Telluric contamination arises from a multiplicative contribution to the science observations from the Earth's atmosphere and is present in all NIR spectroscopic observations.
There are several methods used to correct for telluric contamination that either create a model of the atmospheric conditions~\citep[e.g. ESO's Molecfit tool,][]{2015A&A...576A..77S,2015A&A...576A..78K} or rely on the observation of a  telluric standard star~\citep[e.g.][]{2016MNRAS.458.3968P,2018MNRAS.473.2020L}.

The recommendation from the EMIR instrument team is to observe a telluric standard star for most science cases.
In long-slit mode this can be done with the same observational set-up as the science targets.
In multi-object mode a dedicated slit configuration is used, which is designed to cover the full wavelength range of EMIR with three slitlets placed in the central and two side positions (see below for more details on how slit position changes wavelength coverage).
With this observational set-up telluric contamination can be eliminated on the <5\% level, depending on the signal-to-noise ratio of the standard star.

\section{The CSU design process and configuration tool.}
\label{sec:csu}
The technique of multi-object spectroscopy permits the simultaneous acquisition of many object spectra in a single observation. As previously mentioned, EMIR allows a maximum of 53 objects to be placed in different locations of the FOV. Given the geometry of the CSU (see Fig. \ref{fcsu}), each pair of bars defines an area in the FOV in which only a single object can be observed. Another restriction that the CSU imposes on the observation is that the orientation of each slit in sky is equal for all the slits in the pattern. Any orientation can be set  by varying the instrument position angle, but it is the same for the whole set.

There are two main aspects that the user astronomer has to take into account at the time of building a multi-slit mask at EMIR: the astrometrical precision of the target coordinates and the accuracy of the mask pointing in the sky. Depending on the choice of slit width, the absolute astrometrical precision needed will vary. In general terms, it is advisable to use target object astrometry with at least an accuracy of one-third of the slit width. Gaia and HST coordinates are the recommended input catalogues, but UKIDSS has also been proven to work with mid-size slit widths, between 0.8\arcsec\ and 1.2\arcsec.

The Optimizer Slit Positioner (OSP) tool was specifically designed to cope with all the tasks associated with building the configuration file that will be input into the instrument control software at the GTC to form the desired pattern in the CSU. A detailed description of this tool and additional auxiliary scripts can be found at the EMIR web page \footnote{\url{http://research.iac.es/proyecto/emir/pages/observing-with-emir/observing-tools/osp.php}}. Figure \ref{fosp} shows a picture of a mask being built with the OSP overlaid in a UKIDSS sky image.

For the purpose of this paper, it is sufficient to comment that the OSP takes a target list provided by the user or uploaded from several catalogue servers; reads a fits file to be used as a background image, also supplied by the user or uploaded from servers; and places a CSU mask pattern on it. The user can now displace and rotate the pattern and assign slits of the desired widths to targets in the list. In addition, slits can be joined together to form a longer slit or located in any position on their own sliding zone without being assigned to a source.

The correct positioning of the multi-slit mask on the sky is critical for the success of the observations. This positioning is achieved by the combination of a suitable set of fiducial target objects already assigned to slits in the mask during the designing process and by a dedicated observing strategy implemented in the instrument control software at the GTC. Depending on the number of fiducial stars, their brightness and distribution over the mask, a precision of the order of   0.05\arcsec\ or better can be achieved. It should be noted that the final specific precision in the location of a given target with respect to its assigned slit depends on the quality of the absolute positioning on the sky and on the relative accuracy between the astrometry of the fiducial stars and that of the target objects, which are not always taken from the same catalogues. 

\begin{figure}
        \centering
        \includegraphics[width=\columnwidth]{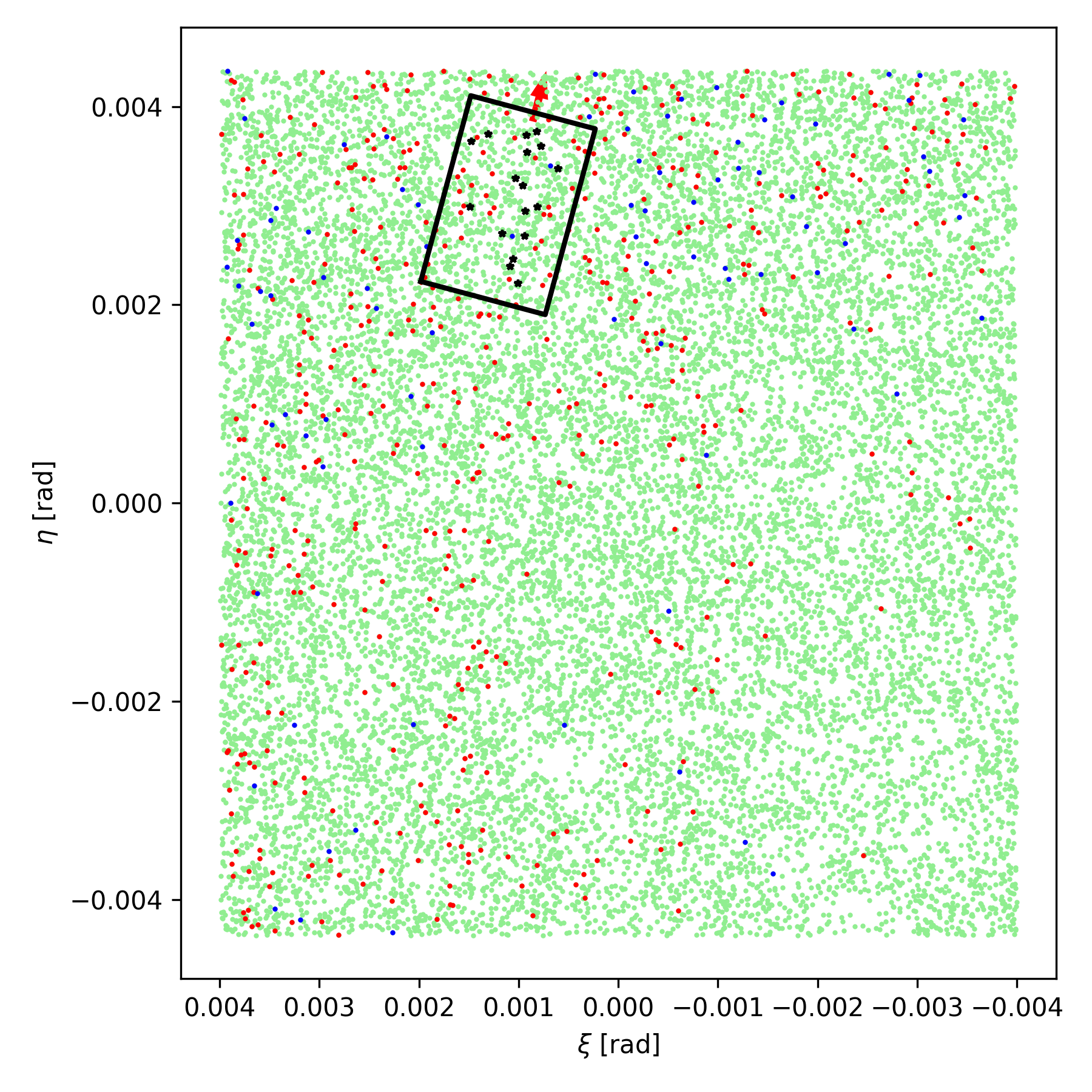}
        \caption{Sketch of an optimum mask positioned on the field. The red arrow shows the orientation of the CSU and defines the position angle of the instrument. The green dots are field objects, the red dots are objects of interest, and the blue dots are bright objects that can be used as fiducial objects for pointing. Inside the mask the selected objects, either red or blue, are represented as black stars. The axes are in standard coordinates}
        \label{fmask1}
\end{figure}

A final remark on the mask design, which  is based on the observation strategy decided by the user astronomer. In the NIR it is a common practice to use a nodding or beam switching scheme to facilitate the measure and removal of the sky background, quite often orders of magnitude brighter than the object. The nodding throw in relation with a single slit length, $\sim 7.2\arcsec$, has to be taken into account at the time of assigning a single slit or a multi-slit to a given target. The presence of companion objects in the slit that can fall in the same pixels as the object of interest or in a neighbour's pixels, either in the initial pointing or in the nodded position, also deserves   consideration at the time of the mask design.

\section{The object catalogue selection tool.}
\label{sec:cat}
The process of designing a multi-object mask, briefly outlined in Sect. \ref{sec:csu}, becomes cumbersome when the object density in the input catalogue is high. The optimal selection of the mask centring is not an easy task; it combines the identification of the best location in the field with the need to avoid contaminating neighbours that should not be present in assigned slits and the need for sufficient fiducial bright objects for proper pointing. In addition, target objects must not be located close to each end of the slit to avoid light loss;  extra adjacent slits must also be allowed for targets depending upon nodding throw. Both positions, initial and nodded, must be free of objects other than the target within a given distance. However, there is an important constraint when placing the targets in a mask that is frequently forgotten; the spectral range covered by each spectrum in a multi-slit mask depends, for a selected spectral band, on the position of the slit in the mask. The image of the slit on the detector marks the central wavelength of the spectrum, which is the same for all the spectra recorded in the mask. Hence, each spectrum spans to both sides from the central position until the detector limit or the wavelength cutoff of the band filter is reached, whichever happens first. However, the user must identify prior to starting the mask design the permitted location of targets in the mask depending on what spectral features are of interest.

\begin{figure}
        \centering
        \includegraphics[width=\columnwidth]{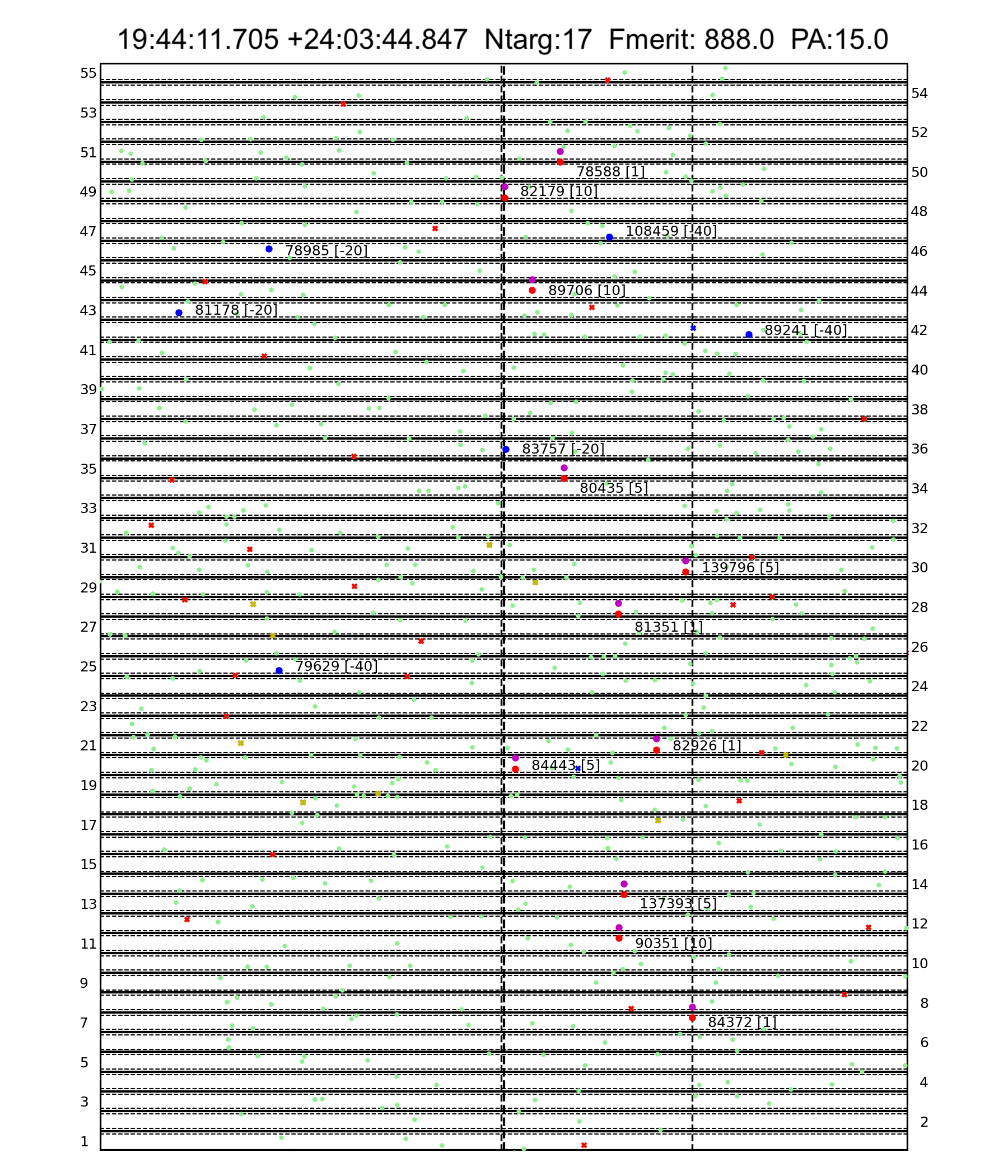}
        \caption{Details of the example mask configuration shown in fig. \ref{fmask1}. Slits are numbered on each side of the mask. The thick vertical dashed line indicates the centre of each slit zone, and the two vertical thin dashed lines limit the area of the mask where the assigned objects must be located. The thick solid horizontal lines are the borders of each slit, and the two thin dashed horizontal lines within each slit limit the effective slit height.  The assigned objects are labelled with the catalogue number and the priority in square brackets. Assigned target objects are represented by a pair of solid circles, red for the initial position and magenta for the nodded one. Assigned fiducial stars are plotted as blue circles. Unassigned targets or fiducial objects are plotted as crosses of the corresponding colour, red for the targets and blue for fiducial stars. Field objects are green circles. There are also yellow crosses representing target objects that cannot be assigned since they have contaminating neighbours in either of the two nodding positions. See text for details.} 
        \label{fmask2}
\end{figure}

We   designed a tool (see \citeads{SALAZARGONZALEZ2021102392} for details) that takes care of most of the aspects that have to be taken into account in a mask design. This is particularly useful in high-density fields, where the large number of individual constraints makes it virtually impossible to find the optimum solution or even a good option for the final mask. The process starts by building an input catalogue on which the objects are graded with different priorities. The tool then makes use of these priorities to find the optimum mask, which is the one that has the largest figure of merit with full compliance of all the user defined constraints. In what follows we   explain   this tool using the GALEP project \citepads{2010ASSP...14..411G} as a pattern. The steps in the mask design can be followed by any user interested in the multi-object spectroscopy mode of EMIR, whether  the mask is designed with the use of the automatic tool or by hand.

GALEP  is an ambitious project that exploits EMIR to obtain near IR spectroscopy of many thousands of Galaxy sources, mainly located in the inner regions. These are selected from their position on IR colour--magnitude diagrams (CMDs) and include disc, bar bulge and ring sources. The principal aim is to accurately classify the sources to provide a better understanding of, particularly, the redder parts of the infrared CMD. GALEP extracts its source candidates from the UKIDSS database \citepads{2007MNRAS.379.1599L}, while several additional infrared catalogues are used for auxiliary information. Most of the sources that GALEP intended to measure lie within or very close to the Galactic plane and towards the internal Galaxy. Hence, most of the fields are of very high source density, which   imposes the use of automatic selection for designing the multi-slit masks.

The first step in the design to automatise it is the translation to numeric values of the scientific merit of the candidates. GALEP selects its candidates to be mostly red giants, of types K and M, and with heliocentric distance between 7 and 9 kpc, while the precise range varies with the location of the field.  We use a model of infrared stellar galactic distribution \citepads{1992ApJS...83..111W}. Using this model,   traces of red giants in the CMD of each field can be located. Then the candidates are selected by their proximity to the theoretical trace and their infrared magnitudes that matches with the selected distance range. The two parameters of proximity and magnitude are combined to derive a numerical priority figure for a given target. In the mask design tool, the figure of merit of each mask is defined as just the sum of the priorities, so several trials should be performed to refine the priority steps before achieving the desired result.

Two important input parameters that the user has to set are the nod throw and the dead space at the two ends of the slit length. The first   is related to the expected size of the target image in the detector that, for the case of point-like sources, is dictated by the actual seeing at the time of the observations. For the GTC at Observatorio del Roque de los Muchachos (La Palma, Spain) a good figure for best seeing in the NIR is 0.6\arcsec, while a mid-range value is around 1\arcsec. It is safe to assume that the  image point spread function width is 3 to 4 times the seeing. So typical values for nodding throw in the case of stellar objects range between 3\arcsec\  and 5\arcsec. The second parameter, the dead space, fixes the effective slit length. The nodding and jittering of the telescope during the observations cause the position of the source in the slit to vary a bit, so it is a good practice not to locate a source very close to a slit end to avoid light loss. A dead space of 1\arcsec\ is normally sufficient. This limitation only applies to the ends of the slit, whether formed by a single pair of bars or by several adjacent pairs.

Finally, the user has to decide    the avoidance zone around targets for which there should be no other object in the catalogue. This is to avoid contamination from neighbours. This avoidance has to be established around the two nodding positions of each potential target. Since the orientation of the mask in the sky and the direction of the nodding are not known a priori, this restriction is quite difficult to handle and has to be verified several times during the mask design, if done manually, to avoid mistakes.

Fortunately, the tool for optimal mask design copes with all the above-mentioned restrictions, and offers a good to optimal result. As explained in \citeads{SALAZARGONZALEZ2021102392}, the tool is based on a non-linear solver for mathematical optimisation problem that delivers a mask with the maximum sum of priorities of all the objects that are assigned to it and comply with the limitation described above. Since finding the optimal solution of the problem could be quite time consuming, it is also possible to limit the CPU time and/or set a value for the figure of merit that, if achieved, stops the process, which means that in these cases the solution might not be the optimum one, but will always satisfy the user's requirements. Figure \ref{fmask1} shows an example of the best mask found by the tool in a UKIDSS field of $0.5\degr\times0.5\degr$ located in the Galactic plane at longitude $\sim60\degr$. For this run we specified 4\arcsec\ of nod throw, 1\arcsec\ of slit dead space, and 3\arcsec\ of avoidance around targets. It is evident from the figure that it would have been virtually impossible, or at least very difficult, to find such a good mask by hand. In fig. \ref{fmask2} the details of the mask can be seen.  At the  top of the plot  the mask centre coordinates are given; the number of objects,  both targets and fiducial stars,  included
in the mask; the figure of merit of this mask (i.e. the sum of the weighted priorities of all the targets); and the position angle of the instrument to position this mask on the sky in the desired orientation. Object priorities are positive for target candidates, 1 being the highest priority, and negative for fiducial objects.  For the fiducial stars there are no provisions for contaminating neighbours nor for the location of the nod position, as they are only used for pointing the mask on the sky.

Figure \ref{fmask2} shows several details of interest for the mask design. First, all the assigned target objects are within the area of interest for the wavelength range whose borders are the two thin vertical dashed lines. As mentioned before, this is extremely important in order to get the proper scientific return from the observations. In fig. \ref{fmask2} all the objects share the same area of interest, but the tool can cope with different areas for different objects. Second, all the target's position pairs lie within the permitted length of the slits (i.e. inside the area limited by the thin horizontal dashed lines). This area can be within a single slitlet,  for instance for object 84443 in slit 20 or object  81351 in slit 28, or can lie in two slitlets,  for example object 90351 in slits 11 and 12 or object 78588 in slits 50 and 51. Finally, it should be checked whether the problem of contaminating neighbours has been avoided.

The above brief description of the mask design process shows a series of steps that must be followed to get the maximum return of an EMIR MOS observation. The automatic tool is available upon request; however, it is  not yet very user friendly.
 
\section{PyEmir, the EMIR data reduction pipeline}
\label{sec:pyemir}

Observing in the NIR, particularly in spectroscopic mode, is a bit tricky and quite often the raw data are plagued by instrumental and background signatures. The removal of all these features is a tedious task and, in most cases, beyond   the reach of the user astronomer who is not fully aware of the instrument specificities. Like most   modern instruments, EMIR   has   its own data reduction pipeline (DRP, PyEmir  \footnote{\url{https://github.com/guaix-ucm/pyemir}}, \citeads{2019hsax.conf..605C}, \citeads{2019ASPC..521..232P}, \citeads{2019ASPC..523..317C}), written in Python, that can cope with the large majority of the different aspects of data reduction. The basic idea of the DRP is to treat the raw data up to a level from  which the scientific analysis can proceed.

\begin{figure}
        \centering
        \includegraphics[width=\columnwidth]{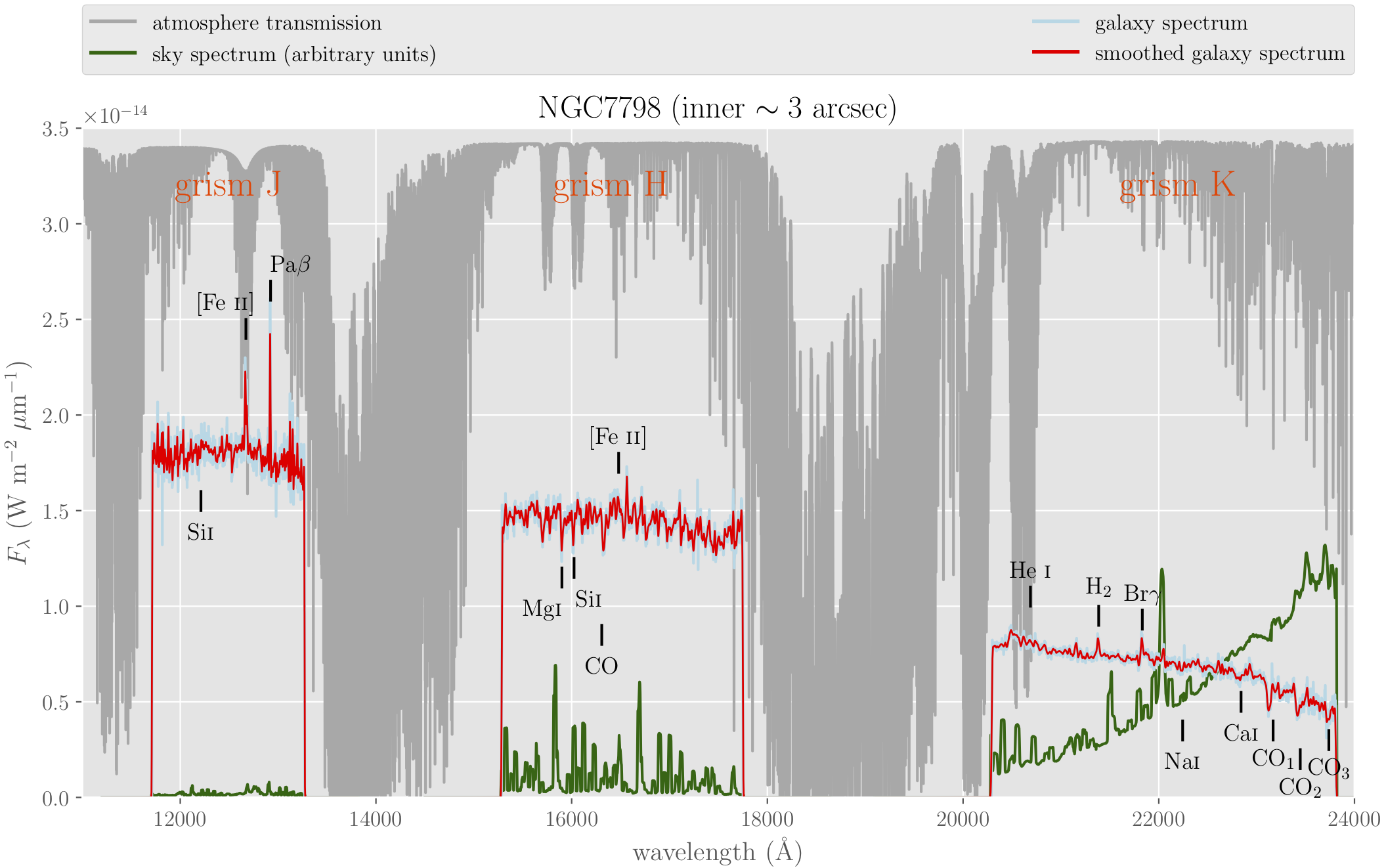}
        \caption{Central spectrum (inner $\sim 3$ arcsec) of the galaxy NGC7798, observed during the commissioning time of EMIR (July 2016). This plot combines the flux-calibrated spectra in $J, H,$ and $K$ (blue curve; smoothed version in red), together with the sky spectrum in arbitrary units (green) and the atmospheric transmission (light grey). Some conspicuous emission lines in the galaxy spectra are labelled. } 
        \label{fpyemir}
\end{figure}

Of particular use is the spectroscopic part of PyEmir, which, apart from the removal of instrumental features,  extracts the single spectra and performs a careful rectification of the initially curved spectra in the detector image and subsequent wavelength calibration. Details can be found at the PyEmir website https://pyemir.readthedocs.io/en/latest/index.html,  including two complete tutorials, one dedicated to the combination of dithered exposures (imaging mode) and one specific to MOS.  As an additional aid, links to some Jupyter notebooks are also included to illustrate the production of a flux-calibrated spectrum of the galaxy NGC7798\footnote{\url{https://guaix-ucm.github.io/pyemir-tutorials/tutorial_mos/ngc7798.html}}, obtained during the commissioning time (see fig. \ref{fpyemir}). 

\section{Scientific results and instrument performance}
\label{sec:sci}
In this section we present a brief overview of a selection of scientific projects that are focused on the spectroscopic exploitation of EMIR.

\subsection{GALEP}
\label{ssec:galep}
\begin{figure}
        \centering
        \includegraphics[width=\columnwidth]{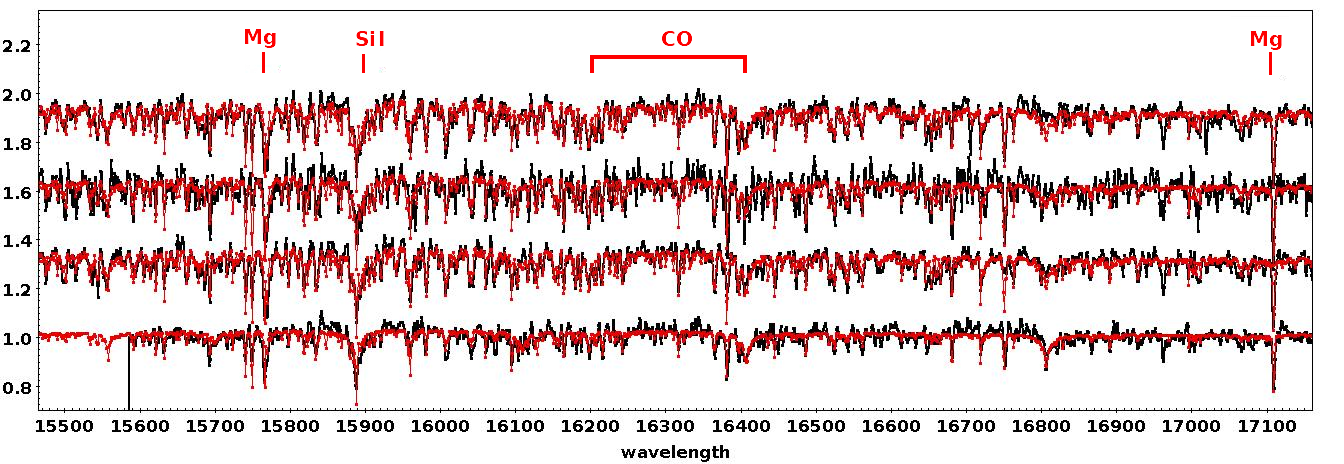}
        \includegraphics[width=\columnwidth]{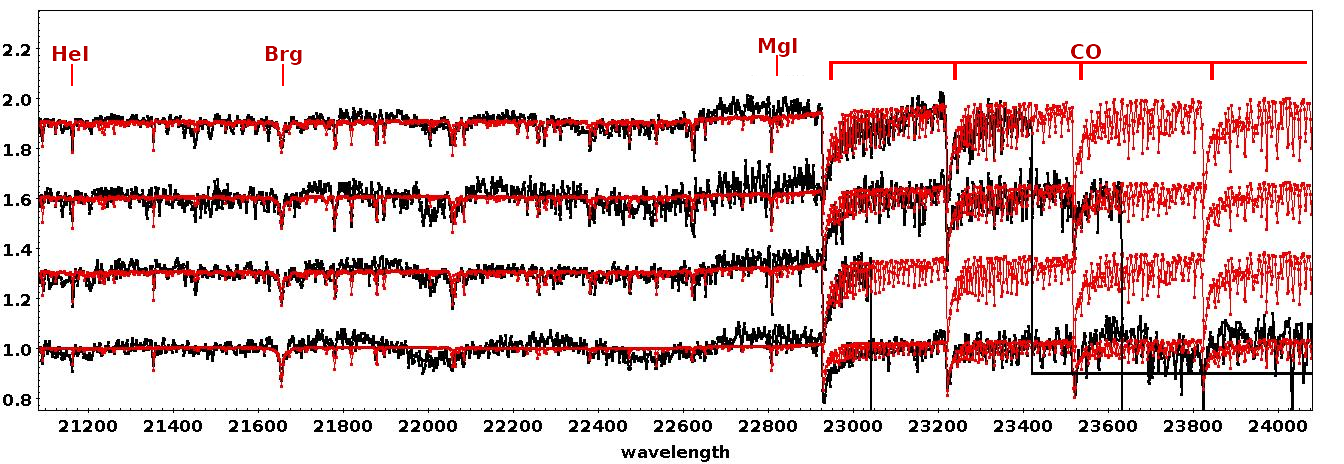}
        \caption{Example of spectra in the H band (upper panel)  and K band (lower panel) of red giants candidates taken with EMIR in MOS mode. The slit width is set as 0.8\arcsec\ for all objects. The red lines above the spectra are the atmospheric models fitted with Ferre \citepads{2006ApJ...636..804A}. The spectra, normalised in flux, are shifted in vertical to fit within the figure. Units on the X-axis are the wavelengths in angstroms.} 
        \label{fgalep}
\end{figure}

Using  EMIR we plan to obtain NIR spectroscopy of several thousand Galactic sources, mainly located in the inner regions. These targets are being selected from their positions on IR colour-- magnitude diagrams and will include disc, bar bulge, and ring sources. The principal aim is to accurately classify the sources to provide a better understanding of, particularly, the redder parts of the infrared CM diagram. Without this information there will remain ambiguities in the interpretation of structures in the inner Galaxy. This project will measure and classify a significant fraction of the sources in specific areas in a way that will avoid the a priori hypotheses for their interpretation. The zones will be spread over a wide area along the plane that can be reachable from the GTC.

The GALEP data will be cross-correlated and compared with other existing databases, like Gaia and LAMOST, on which our group have been working for several years; this will  permit us to explore the Galactic stellar population in a number of ways. The combination of radial velocities, proper motions, parallaxes, and metallicities from Gaia, LAMOST, and APOGEE is opening new opportunities to observe the Galaxy and its morphological components as  time-dependent systems in evolution. GALEP at EMIR will add new capabilities to the already existing databases as it will permit the spectroscopic detection of the stellar population in the hidden central areas of the Galaxy.

Some spectra were   already taken during the commissioning and science verification phases of the EMIR instrument. The data were   reduced using PyEmir (see sec. \ref{sec:pyemir}) and are under analysis using the Ferre code \citepads{2006ApJ...636..804A} to fit the model of atmospheric emission. In figure \ref{fgalep} a few spectra of red giant candidates in the inner Galaxy are shown. The stellar parameters derived with the model range in effective temperature from 4500K to 4900K, in \textit{logg} from 2.4 to 3.1, and in metallicity from -0.3 to 0.0.

\subsection{Searching for obscured massive clusters: MASGOMAS}

The MAssive Stars in Galactic Obscured MAssive clusterS (MASGOMAS) is a survey dedicated to searching for massive stars in the Galactic plane by using photometric cuts in 2MASS \citep{skrutskie06} $K_S$ magnitude, $(J-K_S)$ colour, and $Q_{IR}$ pseudo-colour, and by looking for stellar overdensities after the application of the cuts. The project has been presented  elsewhere, as well as some new clusters observed with LIRIS at the  William Herschel Telescope: MASGOMAS-1 (a very massive cluster candidate hosting OB-type stars and RSG; \citealt{ramirezalegria12}), MASGOMAS-4 (showing two nearby cores; \citealt{ramirezalegria14}), and MASGOMAS-6 (with an alignment along the line of view of two clusters at different distances; \citealt{ramirezalegria18}).

\begin{figure}
    \centering
    \includegraphics[width=6.8 cm]{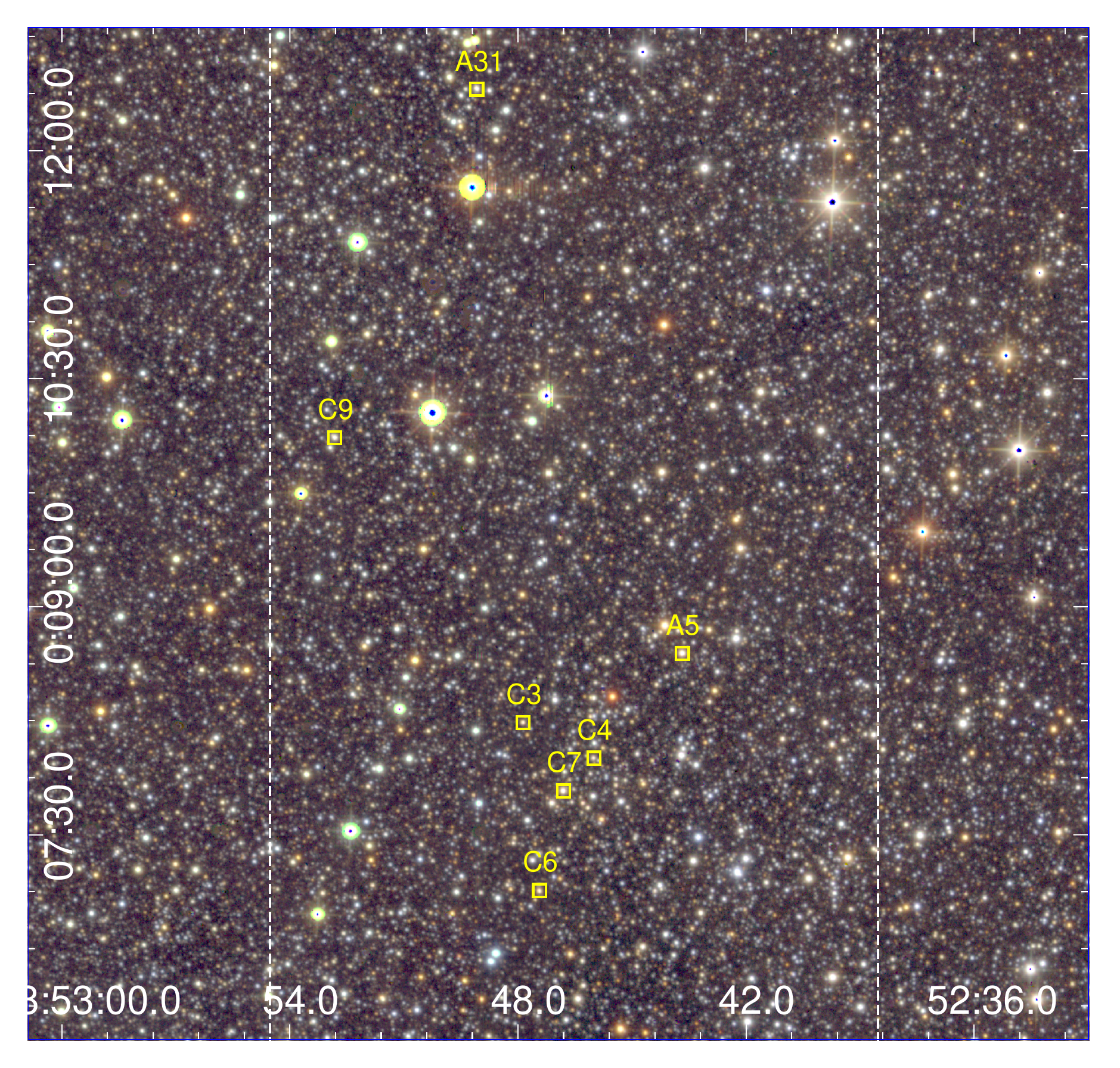}
    \includegraphics[width=2.1 cm]{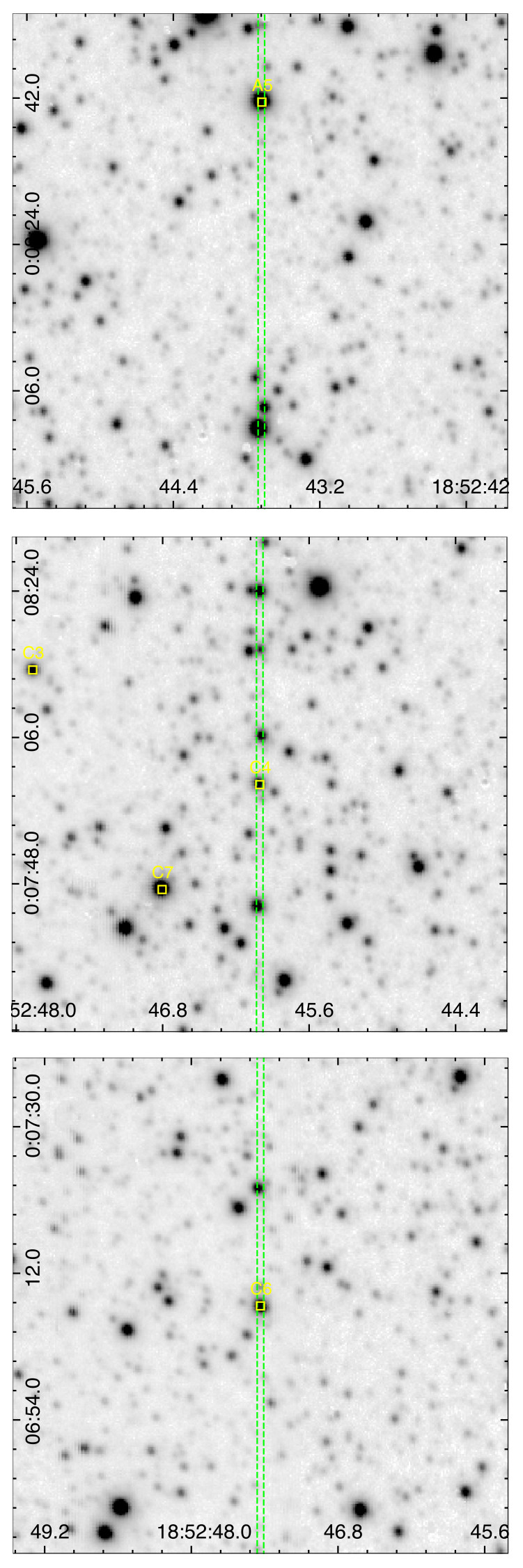}
    \caption{Preparing the observation of the MASGOMAS-10 cluster candidate with EMIR. (Left) False colour UKIDSS-GPS image of MASGOMAS-10 ($J$: blue, $H$: green, $K_S$: red). The white dashed central rectangle shows EMIR's field of view in spectroscopic mode. The stars observed with EMIR during the spectroscopic campaigns are shown as yellow squares (labelled using the internal ID numbers). North is up, east to the left. (Right) Slit position and orientation is critical to exclude contamination during the ABBA observing patterns. These three images show a 0.8 arcsec slit (green dashed rectangle) on three studied stars, and some possible contaminants on the slit. A careful determination of slit parameters reduces this contamination effect and the ABBA overlapping.}
    \label{fig:masgomas10_FOVEmir}
\end{figure}

Using some modification of the techniques presented in the above-mentioned papers we selected a new candidate, MASGOMAS-10. It was observed with EMIR in long-slit mode during September 2017, with a slit width of 0.8 arcsec that projects on four pixels on the detector. We used an ABBA pattern with nodding distances that varied between 5 and 12 arcsec, depending on the observing field, and checked that the telescope motion did not result in the B image of our target overlapping with the A image of some nearby star. Total exposure times ran from 56 to 400 seconds in the $H$ band and from 32 to 240 seconds in the $K$ band. The A0\,V star HIP 92\,711 was used as a standard star to eliminate the telluric lines. Seven candidates of hot massive stars were selected for the observation and are currently being analysed. Figure~\ref{fig:masgomas10_FOVEmir} shows an image of the candidate cluster with the stars observed by EMIR indicated with squares. Table~\ref{tab:masgomas} gives the list of observed stars, with their magnitudes and provisional classifications.

\begin{table*}
    \centering
    \begin{tabular}{r|c|c|c|c|c|c|c}
    ID &  RA (J2000) & DEC (J2000) & $H-$mag & $K-$mag & t$_{exp}$ ($H,K$) & S/N ($H,K$)& Spec. Type\\
    \hline
     A31 & 18:52:49.10 & +00:12:24.2 & 11.66 & 10.89 &  144, -- & 66, -- & O9\,V \\
     A5  & 18:52:43.69 & +00:08:41.5 & 10.68 & 9.61  &   56, 32   & 96, 114 & WN\,8  \\
     C3  & 18:52:47.88 & +00:08:14.3 & 12.83 & 11.86 &  400, 240  & 58, 86 & O8\,V \\
     C4  & 18:52:46.02 & +00:08:00.2 & 12.68 & 11.84 &  360, 240  & 86, 116 & O8\,V \\
     C6  & 18:52:47.45 & +00:07:08.0 & 12.31 & 11.18 &  240, 120  & 90, 86 & B0\,V \\
     C7  & 18:52:46.82 & +00:07:47.3 & 11.00 & 10.04 &  -- , 56   & --, 98 & O8\,III \\
     C9  & 18:52:52.84 & +00:10:06.7 & 11.47 & 10.66 &  120, 96   & 94, 90 & O8\,III   \\
    \end{tabular}
    \caption{Stars observed with EMIR in the candidate cluster MASGOMAS-10. Positions and magnitudes are from the 2MASS Catalog. Exposure times are given in seconds. Signal-to-noise ratio is given by resolution element and was measured around 1.60 $\mu$m in $H$ and 2.13 $\mu$m in $K$.}
    \label{tab:masgomas}
\end{table*}

\begin{figure}
    \centering
    \includegraphics[width=\columnwidth]{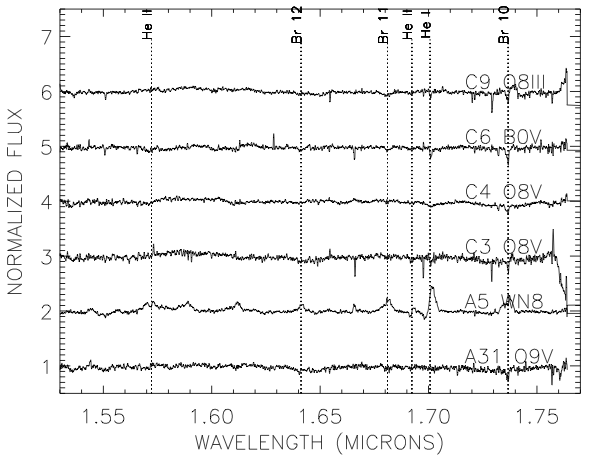}
    \includegraphics[width=\columnwidth]{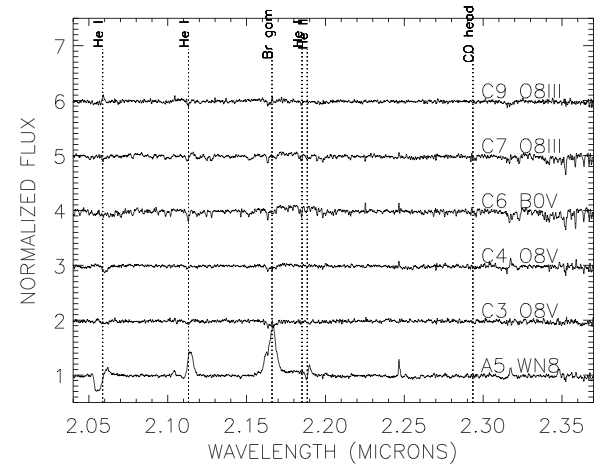}
    \caption{Observed $H$- and $K$-band spectra, top and bottom panels respectively, of stars in the candidate massive cluster MASGOMAS-10 obtained with EMIR}
    \label{fig:masgomassp}
\end{figure}

Figure~\ref{fig:masgomassp} shows the $H$ and $K$ observations in the observed slits. We note that not all stars could be observed in both spectral bands. In addition to the candidate targets other stars (not shown in the figures) appeared in the slits. They turned out to be red stars, as expected from the target selection procedures and evidenced by the strong CO molecular bands starting at 2.29 $\mu$m, which are not seen in the hot stars. The contamination (which did not influence our observations) was particularly large in slit number 6, with up to seven contaminants whose spectra could be extracted. Observations with EMIR have thus allowed us to confirm the presence of obscured massive hot stars in a small region of the sky in just over half an hour of on-target exposure times.

\subsection{EMIR spectroscopy of star-forming galaxies at redshift $z\sim0.9$}
\label{ssec:goya}
EMIR is particularly useful for obtaining rest-frame optical or UV spectra of high-redshift galaxies   shifted to the NIR by cosmological expansion. 
The GOYA survey was designed to exploit this capability to address key questions of intermediate- and high-redshift galaxy evolution. GOYA encompasses sub-projects to study intermediate-redshift, low-mass star-forming galaxies, intermediate-redshift massive galaxies, and galaxies in the epoch of reionization. 
GOYA aims to survey several cosmological fields using the multiple capabilities of the EMIR MOS mode. The Extended Groth Strip (EGS) was the field selected for the first batch of observations. These projects plan to start obtaining data once the commissioning of the Hawaii-2RG detector is completed. 

We present here the first results of a pilot spectroscopic study of a sample of low-mass star-forming galaxies at intermediate redshifts ($0.7 < z < 1.5$), specifically selected to study the formation redshift and the cosmic role of dwarf galaxies. The high surface density of star-forming galaxies qualifies them as ideal fillers for the GOYA survey MOS masks, guarantying the coverage of a minimum sample. 

\begin{figure}
    \centering
    \includegraphics[width=\columnwidth]{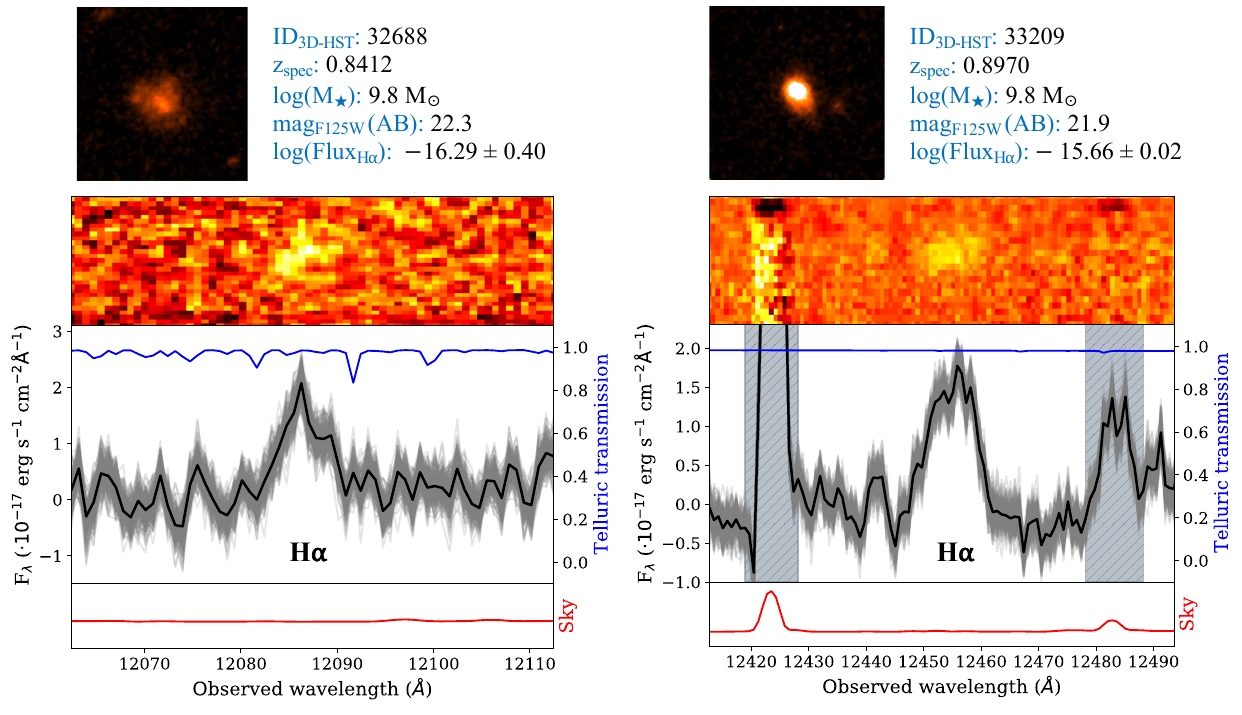}
    \caption{Example of H$\alpha$ emission lines for faint star-forming galaxies at z$\sim$1. See details in text.}
    \label{goya1}
\end{figure}

In figure \ref{goya1} we present two star-forming galaxies at intermediate redshift that show H$\alpha$ emission in EMIR long-exposure MOS masks. The panels are from images obtained with the HST filter WFC3/F125W, size $5\arcsec \times 5\arcsec$, and   orientated north up and east left. The physical parameters are from the 3D-HST catalogue,  and the H$\alpha$ fluxes are those estimated by the 3D-HST team from the HST G141 grism low-resolution spectra (\citeads{2014ApJS..214...24S}). Two-dimensional spectra centred at H$\alpha$ are shown. The emission line is clearly visible, but the continuum can only be estimated. The profiles show the redshifted H$\alpha$ emission line, telluric transmission, and sky brightness. From the analysis of the emission lines, we obtained the following values: (a) $ID_{3D-HST}=32688,\ z_{spec}=0.84162 \pm 0.00004,\ log(Flux_{H\alpha})=-15.99 \pm 0.04$; (b) $ID_{3D-HST}=33209, z_{spec}=0.89778 \pm 0.00003,\ log(Flux_{H\alpha})=-15.84 \pm 0.02$. 

These preliminary detections in the $J$ band show the H$\alpha$ emission of star-forming galaxies, in good agreement with the H$\alpha$ fluxes estimated from the HST G141 grism by the 3D-HST survey.

\subsection{Some prospects on the use of EMIR to target the reionization era}
\label{ssec:goya_highz}
One of the main goals of the GOYA survey with EMIR was precisely the identification and study of the first generation of galaxies formed in the universe. This area is being massively explored by JWST. However, EMIR remains an interesting and complementary instrument in the JWST era, due to its multi-plex capability over a wide FOV together with a relatively high spectral resolution, which  allow   astronomers to  resolve   emission-line profiles, such as   the Lyman$\alpha$ lines (see e.g. \citeads{2021MNRAS.500..558M}). The spectral resolution achieved by EMIR is at least a factor of two higher than for the highest resolution mode of NIRSpec, and the FOV available for spectroscopy is more than two times larger (\citeads{2022A&A...661A..80J}). A relatively large FOV is important to derive statistically significant results when studying the abundance and global properties of z$\sim6-12$ galaxies based on spectroscopic samples. EMIR is also ideally suited to systematically explore the properties of reionization bubbles, and the ionising sources residing inside these bubbles (see e.g. \citeads{2022A&A...662A.115C}), a key study to understand the reionization process that is still limited by observational capabilities. 

Due to the technical problems mentioned above, our first attempt to confirm the redshift of the z$=$8.68 galaxy reported by  \citeads{2015ApJ...810L..12Z} by the re-detection of the Ly$\alpha$ line expected in the $J$ band was unsuccessful, despite a comparable exposure time with respect to the original observations with the MOSFIRE/Keck telescope, and good observing conditions. New observations are expected in the near future once these technical problems are   fixed, targeting both blank fields and strong-lensing clusters. Photometric samples are being constructed for this survey, specifically optimised for EMIR.  

\section{Concluding remarks}
\label{sec:conclu}
In this paper we have summarised the main capabilities of the EMIR instrument from an astronomical standpoint. We have described the techniques to fully exploit the multi-object mode of EMIR, which is the one that deserves specific attention from the users as it is not as straightforward to  use as the standard image and long-slit spectroscopy observing modes. 

The references in the bibliography can provide the interested reader full details of the EMIR instrument, its design, specifications and main instrumental features.
We have shown that EMIR is a powerful instrument that is currently at the disposal of the astronomical community at the  largest optical and infrared telescope in the world, the GTC in the Spanish island of La Palma.

\begin{acknowledgements} 
EMIR has been funded by GRANTECAN S.L. via a procurement contract; by the Spanish funding agency grants AYA2001-1656, AYA2002-10256-E, FIT-020100-2003-587, AYA2003-01186, AYA2006-15698-C02-01, AYA2009-06972, AYA2012-33211, AYA2015-63650-P and AYA2015-70498-C2-1-R; and by the Canarian funding agency grant ACIISI-PI 2008/226. Part of this work was supported by AYA2018-RTI-096188-B-I00 and the French CNRS, the Aix-Marseille University, the French Programme National de Cosmologie et Galaxies (PNCG) of CNRS/INSU with INP and IN2P3, co-funded by CEA and CNES. 
\end{acknowledgements}

\bibliographystyle{aa}
\bibliography{emir_aa.bib}

\begin{thebibliography}{28}
\expandafter\ifx\csname natexlab\endcsname\relax\def\natexlab#1{#1}\fi

\bibitem[{{Allende Prieto} {et~al.}(2006){Allende Prieto}, {Beers}, {Wilhelm},
  {Newberg}, {Rockosi}, {Yanny}, \& {Lee}}]{2006ApJ...636..804A}
{Allende Prieto}, C., {Beers}, T.~C., {Wilhelm}, R., {et~al.} 2006, \apj, 636,
  804

\bibitem[{{Balcells}(2003)}]{2003RMxAC..16...69B}
{Balcells}, M. 2003, in Revista Mexicana de Astronomia y Astrofisica Conference
  Series, Vol.~16, Revista Mexicana de Astronomia y Astrofisica Conference
  Series, ed. J.~M. {Rodriguez Espinoza}, F.~{Garzon Lopez}, \& V.~{Melo
  Martin}, 69--72

\bibitem[{{Cardiel} {et~al.}(2019{\natexlab{a}}){Cardiel}, {Pascual},
  {Gallego}, {Cabello}, {Garz{\'o}n}, {Balcells}, {Castro-Rodr{\'\i}guez},
  {Dom{\'\i}nguez-Palmero}, {Hammersley}, {Laporte}, {Patrick}, {Pell{\'o}},
  {Prieto}, \& {Streblyanska}}]{2019ASPC..523..317C}
{Cardiel}, N., {Pascual}, S., {Gallego}, J., {et~al.} 2019{\natexlab{a}}, in
  Astronomical Society of the Pacific Conference Series, Vol. 523, Astronomical
  Data Analysis Software and Systems XXVII, ed. P.~J. {Teuben}, M.~W. {Pound},
  B.~A. {Thomas}, \& E.~M. {Warner}, 317

\bibitem[{{Cardiel} {et~al.}(2019{\natexlab{b}}){Cardiel}, {Pascual},
  {Gallego}, {Cabello}, {Garz{\'o}n}, {Balcells}, {Castro-Rodr{\'\i}guez},
  {Dom{\'\i}nguez-Palmero}, {Hammersley}, {Laporte}, {Patrick}, {Pell{\'o}},
  {Prieto}, \& {Streblyanska}}]{2019hsax.conf..605C}
{Cardiel}, N., {Pascual}, S., {Gallego}, J., {et~al.} 2019{\natexlab{b}}, in
  Highlights on Spanish Astrophysics X, ed. B.~{Montesinos}, A.~{Asensio
  Ramos}, F.~{Buitrago}, R.~{Sch{\"o}del}, E.~{Villaver}, S.~{P{\'e}rez-Hoyos},
  \& I.~{Ord{\'o}{\~n}ez-Etxeberria}, 605--605

\bibitem[{{Castellano} {et~al.}(2022){Castellano}, {Pentericci}, {Cupani},
  {Curtis-Lake}, {Vanzella}, {Amor{\'\i}n}, {Belfiori}, {Calabr{\`o}},
  {Carniani}, {Charlot}, {Chevallard}, {Dayal}, {Dickinson}, {Ferrara},
  {Fontana}, {Giallongo}, {Hutter}, {Merlin}, {Paris}, \&
  {Santini}}]{2022A&A...662A.115C}
{Castellano}, M., {Pentericci}, L., {Cupani}, G., {et~al.} 2022, \aap, 662,
  A115

\bibitem[{{Garz{\'o}n} {et~al.}(2006){Garz{\'o}n}, {Abreu}, {Barrera},
  {Becerril}, {Cair{\'o}s}, {D{\'\i}az}, {Fragoso}, {Gago}, {Grange},
  {Gonz{\'a}lez}, {L{\'o}pez}, {Patr{\'o}n}, {P{\'e}rez}, {Rasilla}, {Redondo},
  {Restrepo}, {Saavedra}, {S{\'a}nchez}, {Tenegi}, \&
  {Vallb{\'e}}}]{2006SPIE.6269E..18G}
{Garz{\'o}n}, F., {Abreu}, D., {Barrera}, S., {et~al.} 2006, in Society of
  Photo-Optical Instrumentation Engineers (SPIE) Conference Series, Vol. 6269,
  Society of Photo-Optical Instrumentation Engineers (SPIE) Conference Series,
  ed. I.~S. {McLean} \& M.~{Iye}, 626918

\bibitem[{{Garz{\'o}n} {et~al.}(2016){Garz{\'o}n}, {Castro}, {Insausti},
  {Manjavacas}, {Miluzio}, {Hammersley}, {Cardiel}, {Pascual},
  {Gonz{\'a}lez-Fern{\'a}ndez}, {Molg{\'o}}, {Barreto}, {Fern{\'a}ndez},
  {Joven}, {L{\'o}pez}, {Mato}, {Moreno}, {N{\'u}{\~n}ez}, {Patr{\'o}n},
  {Rosich}, \& {Vega}}]{2016SPIE.9908E..1JG}
{Garz{\'o}n}, F., {Castro}, N., {Insausti}, M., {et~al.} 2016, in Society of
  Photo-Optical Instrumentation Engineers (SPIE) Conference Series, Vol. 9908,
  Ground-based and Airborne Instrumentation for Astronomy VI, ed. C.~J.
  {Evans}, L.~{Simard}, \& H.~{Takami}, 99081J

\bibitem[{{Garz{\'o}n} {et~al.}(2017){Garz{\'o}n}, {Castro}, {Insausti},
  {Manjavacas}, {Miluzzio}, {Hammersley}, {Cardiel}, {Pascual},
  {Gonz{\'a}lez-Fern{\'a}ndez}, {Molg{\'o}}, {Barreto}, {Fern{\'a}ndez},
  {Joven}, {L{\'o}pez}, {Mato}, {Moreno}, {N{\'u}nez}, {Patr{\'o}n}, {Rosich},
  \& {Vega}}]{2017hsa9.conf..652G}
{Garz{\'o}n}, F., {Castro}, N., {Insausti}, M., {et~al.} 2017, in Highlights on
  Spanish Astrophysics IX, ed. S.~{Arribas}, A.~{Alonso-Herrero},
  F.~{Figueras}, C.~{Hern{\'a}ndez-Monteagudo}, A.~{S{\'a}nchez-Lavega}, \&
  S.~{P{\'e}rez-Hoyos}, 652--659

\bibitem[{{Garz{\'o}n} \& {EMIR Team}(2016)}]{2016ASPC..507..297G}
{Garz{\'o}n}, F. \& {EMIR Team}. 2016, in Astronomical Society of the Pacific
  Conference Series, Vol. 507, Multi-Object Spectroscopy in the Next Decade:
  Big Questions, Large Surveys, and Wide Fields, ed. I.~{Skillen},
  M.~{Balcells}, \& S.~{Trager}, 297

\bibitem[{{Garz{\'o}n} {et~al.}(2010){Garz{\'o}n}, {Hammersley},
  {Gonz{\'a}lez}, \& {Cabrera}}]{2010ASSP...14..411G}
{Garz{\'o}n}, F., {Hammersley}, P.~L., {Gonz{\'a}lez}, C., \& {Cabrera}, A.
  2010, in Astrophysics and Space Science Proceedings, Vol.~14, Highlights of
  Spanish Astrophysics V, 411

\bibitem[{{Guzman}(2003)}]{2003RMxAC..16..209G}
{Guzman}, R. 2003, in Revista Mexicana de Astronomia y Astrofisica Conference
  Series, Vol.~16, Revista Mexicana de Astronomia y Astrofisica Conference
  Series, ed. J.~M. {Rodriguez Espinoza}, F.~{Garzon Lopez}, \& V.~{Melo
  Martin}, 209--212

\bibitem[{{Jakobsen} {et~al.}(2022){Jakobsen}, {Ferruit}, {Alves de Oliveira},
  {Arribas}, {Bagnasco}, {Barho}, {Beck}, {Birkmann}, {B{\"o}ker}, {Bunker},
  {Charlot}, {de Jong}, {de Marchi}, {Ehrenwinkler}, {Falcolini}, {Fels},
  {Franx}, {Franz}, {Funke}, {Giardino}, {Gnata}, {Holota}, {Honnen}, {Jensen},
  {Jentsch}, {Johnson}, {Jollet}, {Karl}, {Kling}, {K{\"o}hler}, {Kolm},
  {Kumari}, {Lander}, {Lemke}, {L{\'o}pez-Caniego}, {L{\"u}tzgendorf},
  {Maiolino}, {Manjavacas}, {Marston}, {Maschmann}, {Maurer}, {Messerschmidt},
  {Moseley}, {Mosner}, {Mott}, {Muzerolle}, {Pirzkal}, {Pittet}, {Plitzke},
  {Posselt}, {Rapp}, {Rauscher}, {Rawle}, {Rix}, {R{\"o}del}, {Rumler},
  {Sabbi}, {Salvignol}, {Schmid}, {Sirianni}, {Smith}, {Strada}, {te Plate},
  {Valenti}, {Wettemann}, {Wiehe}, {Wiesmayer}, {Willott}, {Wright}, {Zeidler},
  \& {Zincke}}]{2022A&A...661A..80J}
{Jakobsen}, P., {Ferruit}, P., {Alves de Oliveira}, C., {et~al.} 2022, \aap,
  661, A80

\bibitem[{{Kausch} {et~al.}(2015){Kausch}, {Noll}, {Smette}, {Kimeswenger},
  {Barden}, {Szyszka}, {Jones}, {Sana}, {Horst}, \&
  {Kerber}}]{2015A&A...576A..78K}
{Kausch}, W., {Noll}, S., {Smette}, A., {et~al.} 2015, \aap, 576, A78

\bibitem[{{Lawrence} {et~al.}(2007){Lawrence}, {Warren}, {Almaini}, {Edge},
  {Hambly}, {Jameson}, {Lucas}, {Casali}, {Adamson}, {Dye}, {Emerson},
  {Foucaud}, {Hewett}, {Hirst}, {Hodgkin}, {Irwin}, {Lodieu}, {McMahon},
  {Simpson}, {Smail}, {Mortlock}, \& {Folger}}]{2007MNRAS.379.1599L}
{Lawrence}, A., {Warren}, S.~J., {Almaini}, O., {et~al.} 2007, \mnras, 379,
  1599

\bibitem[{{Lodieu} {et~al.}(2018){Lodieu}, {Zapatero Osorio}, {B{\'e}jar}, \&
  {Pe{\~n}a Ram{\'\i}rez}}]{2018MNRAS.473.2020L}
{Lodieu}, N., {Zapatero Osorio}, M.~R., {B{\'e}jar}, V.~J.~S., \& {Pe{\~n}a
  Ram{\'\i}rez}, K. 2018, \mnras, 473, 2020

\bibitem[{{Meyer} {et~al.}(2021){Meyer}, {Laporte}, {Ellis}, {Verhamme}, \&
  {Garel}}]{2021MNRAS.500..558M}
{Meyer}, R.~A., {Laporte}, N., {Ellis}, R.~S., {Verhamme}, A., \& {Garel}, T.
  2021, \mnras, 500, 558

\bibitem[{{Pascual} {et~al.}(2019){Pascual}, {Cardiel}, {Garz{\'o}n},
  {Castro-Rodr{\'\i}guez}, {Gonz{\'a}lez-Fern{\'a}ndez}, {Hammersley},
  {Manjavacas}, \& {Miluzio}}]{2019ASPC..521..232P}
{Pascual}, S., {Cardiel}, N., {Garz{\'o}n}, F., {et~al.} 2019, in Astronomical
  Society of the Pacific Conference Series, Vol. 521, Astronomical Data
  Analysis Software and Systems XXVI, ed. M.~{Molinaro}, K.~{Shortridge}, \&
  F.~{Pasian}, 232

\bibitem[{{Patrick} {et~al.}(2016){Patrick}, {Evans}, {Davies}, {Kudritzki},
  {H{\'e}nault-Brunet}, {Bastian}, {Lapenna}, \&
  {Bergemann}}]{2016MNRAS.458.3968P}
{Patrick}, L.~R., {Evans}, C.~J., {Davies}, B., {et~al.} 2016, \mnras, 458,
  3968

\bibitem[{{Ram{\'\i}rez Alegr{\'\i}a} {et~al.}(2018){Ram{\'\i}rez
  Alegr{\'\i}a}, {Herrero}, {R{\"u}bke}, {Mar{\'\i}n-Franch}, {Garc{\'\i}a}, \&
  {Borissova}}]{ramirezalegria18}
{Ram{\'\i}rez Alegr{\'\i}a}, S., {Herrero}, A., {R{\"u}bke}, K., {et~al.} 2018,
  \aap, 614, A116

\bibitem[{{Ram{\'\i}rez Alegr{\'\i}a} {et~al.}(2012){Ram{\'\i}rez
  Alegr{\'\i}a}, {Mar{\'\i}n-Franch}, \& {Herrero}}]{ramirezalegria12}
{Ram{\'\i}rez Alegr{\'\i}a}, S., {Mar{\'\i}n-Franch}, A., \& {Herrero}, A.
  2012, \aap, 541, A75

\bibitem[{{Ram{\'\i}rez Alegr{\'\i}a} {et~al.}(2014){Ram{\'\i}rez
  Alegr{\'\i}a}, {Mar{\'\i}n-Franch}, \& {Herrero}}]{ramirezalegria14}
{Ram{\'\i}rez Alegr{\'\i}a}, S., {Mar{\'\i}n-Franch}, A., \& {Herrero}, A.
  2014, \aap, 567, A66

\bibitem[{Salazar-González(2021)}]{SALAZARGONZALEZ2021102392}
Salazar-González, J.-J. 2021, Omega, 103, 102392

\bibitem[{{Skelton} {et~al.}(2014){Skelton}, {Whitaker}, {Momcheva}, {Brammer},
  {van Dokkum}, {Labb{\'e}}, {Franx}, {van der Wel}, {Bezanson}, {Da Cunha},
  {Fumagalli}, {F{\"o}rster Schreiber}, {Kriek}, {Leja}, {Lundgren}, {Magee},
  {Marchesini}, {Maseda}, {Nelson}, {Oesch}, {Pacifici}, {Patel}, {Price},
  {Rix}, {Tal}, {Wake}, \& {Wuyts}}]{2014ApJS..214...24S}
{Skelton}, R.~E., {Whitaker}, K.~E., {Momcheva}, I.~G., {et~al.} 2014, \apjs,
  214, 24

\bibitem[{{Skrutskie} {et~al.}(2006){Skrutskie}, {Cutri}, {Stiening},
  {Weinberg}, {Schneider}, {Carpenter}, {Beichman}, {Capps}, {Chester},
  {Elias}, {Huchra}, {Liebert}, {Lonsdale}, {Monet}, {Price}, {Seitzer},
  {Jarrett}, {Kirkpatrick}, {Gizis}, {Howard}, {Evans}, {Fowler}, {Fullmer},
  {Hurt}, {Light}, {Kopan}, {Marsh}, {McCallon}, {Tam}, {Van Dyk}, \&
  {Wheelock}}]{skrutskie06}
{Skrutskie}, M.~F., {Cutri}, R.~M., {Stiening}, R., {et~al.} 2006, \aj, 131,
  1163

\bibitem[{{Smette} {et~al.}(2015){Smette}, {Sana}, {Noll}, {Horst}, {Kausch},
  {Kimeswenger}, {Barden}, {Szyszka}, {Jones}, {Gallenne}, {Vinther},
  {Ballester}, \& {Taylor}}]{2015A&A...576A..77S}
{Smette}, A., {Sana}, H., {Noll}, S., {et~al.} 2015, \aap, 576, A77

\bibitem[{{Teuwen} {et~al.}(2012){Teuwen}, {Janssen}, {Casalta}, \& {Garz{\'o}n
  Lopez}}]{2012SPIE.8446E..5NT}
{Teuwen}, M., {Janssen}, H., {Casalta}, J.~M., \& {Garz{\'o}n Lopez}, F. 2012,
  in Society of Photo-Optical Instrumentation Engineers (SPIE) Conference
  Series, Vol. 8446, Ground-based and Airborne Instrumentation for Astronomy
  IV, ed. I.~S. {McLean}, S.~K. {Ramsay}, \& H.~{Takami}, 84465N

\bibitem[{{Wainscoat} {et~al.}(1992){Wainscoat}, {Cohen}, {Volk}, {Walker}, \&
  {Schwartz}}]{1992ApJS...83..111W}
{Wainscoat}, R.~J., {Cohen}, M., {Volk}, K., {Walker}, H.~J., \& {Schwartz},
  D.~E. 1992, \apjs, 83, 111

\bibitem[{{Zitrin} {et~al.}(2015){Zitrin}, {Labb{\'e}}, {Belli}, {Bouwens},
  {Ellis}, {Roberts-Borsani}, {Stark}, {Oesch}, \&
  {Smit}}]{2015ApJ...810L..12Z}
{Zitrin}, A., {Labb{\'e}}, I., {Belli}, S., {et~al.} 2015, \apjl, 810, L12

\end{thebibliography}

\end{document}